\documentclass[aps,prd,superscriptaddress,preprintnumbers,nofootinbib,10pt,longbibliography]{revtex4-2}
\usepackage{multirow}
\usepackage{amsmath}
\usepackage{amssymb}
\usepackage[dvipdf,dvips]{graphicx}
\usepackage{color}
\usepackage{hyperref}
\usepackage{url}
\usepackage{slashed}
\usepackage[usenames,dvipsnames]{xcolor}
\usepackage{amsmath,subcaption}
\usepackage{amsfonts}
\usepackage{float}
\usepackage{amssymb}
\usepackage{epsfig}
\usepackage{graphics}
\usepackage{euscript}
\usepackage{slashed}
\usepackage{epstopdf}
\usepackage[utf8]{inputenc}
\allowdisplaybreaks
\usepackage[normalem]{ulem}
\usepackage{pifont}
\usepackage{dsfont}
\usepackage{graphicx}
\usepackage{latexsym,braket}
\usepackage{tikz-feynman}
\usepackage{tikz-cd}
\usepackage{easyReview}
\usepackage{cancel}
\usepackage[normalem]{ulem}
\usepackage{svg}
\usepackage{cleveref}
\usepackage{physics}
\usepackage{mathtools}
\usepackage{bbm}

\newcommand{\p}{\partial}

\makeatletter
\DeclareRobustCommand{\cev}[1]{%
	{\mathpalette\do@cev{#1}}%
}
\newcommand{\do@cev}[2]{%
	\vbox{\offinterlineskip
		\sbox\z@{$\m@th#1 x$}%
		\ialign{##\cr
			\hidewidth\reflectbox{$\m@th#1\vec{}\mkern4mu$}\hidewidth\cr
			\noalign{\kern-\ht\z@}
			$\m@th#1#2$\cr
		}%
	}%
}
\makeatother

\makeatletter
\newlength{\negph@wd}
\DeclareRobustCommand{\negphantom}[1]{%
  \ifmmode
    \mathpalette\negph@math{#1}%
  \else
    \negph@do{#1}%
  \fi
}
\newcommand{\negph@math}[2]{\negph@do{$\m@th#1#2$}}
\newcommand{\negph@do}[1]{%
  \settowidth{\negph@wd}{#1}%
  \hspace*{-\negph@wd}%
}
\makeatother

\renewcommand{\dd}[2][]{\!\mathrm{d}\ifx#1\empty #2\else^{#1} #2\fi\,}
\newcommand{\ddk}[2][]{%
  \!\ifx#1\empty
    \frac{\mathrm{d} #2}{2\pi}%
  \else
    \frac{\mathrm{d}^{#1} #2}{(2\pi)^{#1}}%
  \fi\,%
}

\newcommand{\M}{\mathcal{M}}

\begin{document}

\title{Examples of infinitesimal Gribov copies in the presence of planar boundaries,\\and an emergent boundary gauge invariance in Yang-Mills theory}

\author{David Dudal}
\email{david.dudal@kuleuven.be}
\affiliation{KU Leuven Campus Kortrijk -- Kulak, Department of Physics, Etienne Sabbelaan 53 bus 7657, 8500 Kortrijk, Belgium}

\author{Luigi Rosa}
\email{rosa@na.infn.it}
\affiliation{INFN, Sezione di Napoli, Complesso Universitario di Monte S.~Angelo, Via Cintia Edificio 6, 80126 Naples, Italia}
\affiliation{Dipartimento di Matematica e Applicazioni ``R.~Caccioppoli'', Universit\'{a} di Napoli Federico II, Complesso Universitario di Monte S.~Angelo,  Via Cintia Edificio 6,
80126 Naples, Italia}

\author{Sebbe Stouten}
\email{sebbe.stouten@kuleuven.be}
\affiliation{KU Leuven Campus Kortrijk -- Kulak, Department of Physics, Etienne Sabbelaan 53 bus 7657, 8500 Kortrijk, Belgium}

\begin{abstract}
    First, we explicitly construct zero-modes of the Yang-Mills Faddeev-Popov operator in the presence of two parallel plates, with either perfect electric (PEC) or perfect magnetic (PMC) boundary conditions imposed. This establishes the existence of infinitesimal Gribov copies with finite action, and even finite $L_2$-norm, in Yang-Mills theory with non-trivial interfaces.
    We adapt Henyey's construction, although the boundary configuration imposes extra restrictions on the ansatz for the gauge field.
    Second, we discuss the phenomenon of an emergent boundary gauge invariance for PEC and PMC plates in Yang-Mills theory.
    This shall be important when applying the Gribov-Zwanziger procedure to parallel plate setups, of relevance for non-perturbatively studying the non-Abelian Casimir effect in the future.
\end{abstract}
\maketitle
\date{}

\section{Introduction}

It is well-known that Yang-Mills (YM) theory under the Faddeev-Popov (FP) gauge fixing procedure suffers from Gribov copies: different gauge field configurations on the same gauge orbit that obey the same gauge fixing condition \cite{Gribov78,Singer78}.
In other words, the FP procedure does not fully fix the gauge, and there is some overcounting in the functional integral.
Moreover, since the existence of infinitesimal Gribov copies is equivalent to the existence of zero-modes for the FP operator, the FP determinant can become zero or negative.
Since it is supposed to be positive in order to act as a proper measure, this possibly renders the whole FP procedure invalid, which is especially problematic in the infrared (IR) region.
Lattice QCD simulations have shown important non-perturbative features exactly in the IR region \cite{Bogolubsky:2009dc,Cucchieri:2007md,Cucchieri:2007rg,Cucchieri:2008fc,Oliveira:2012eh,Dudal:2018cli,Maas:2008ri,Schleifenbaum:2004id,Colaco:2024gmt}. In particular, the gluon propagator tends to a non-zero finite value for vanishing momenta, and the coupling constant seems not to encounter a Landau pole.
An effective theory that can efficiently replicate these observations is the Curci-Ferrari model \cite{CurciFerrari76,Tissier10,Tissier11,Pelaez21}: YM theory in Landau gauge, albeit with a gluon mass term added by hand \emph{after} quantization.
Instead of this effective model, a more first-principles alternative is the Gribov-Zwanziger approach \cite{Gribov78,Zwanziger89,Vandersickel11,Dudal:2008sp,Capri16}.
Other non-perturbative techniques such as Schwinger-Dyson equations, the Functional Renormalization Group \cite{Alkofer:2000wg,Aguilar:2008xm,Fischer:2006ub,Fischer:2008uz,Bashir:2012fs,Huber:2020keu,Boucaud:2011ug}, and dynamically massive versions of YM \cite{Comitini:2017zfp,Comitini:2020ozt,Comitini:2023urc} have also been used to study the IR properties of YM and the Gribov problem.

This Gribov problem is not merely academic in nature, as explicit constructions of such Gribov copies have been carried out in e.g.~\cite{SCIUTO1979181,Jackiw:1978, Henyey:1978qd}, and more recently in \cite{Guimaraes:2011sf,Capri:2012ev,Landim:2012eg,Anabalon:2010um,Canfora:2011xd}.
Both the theoretical proof of existence and the explicit constructions are done in empty spacetimes.
However, there are important physical phenomena (such as the Casimir effect) that occur in non-empty spacetimes.
To the best of our knowledge, for such settings, no research has been done on the existence of Gribov copies.
In this work, we aim to close this gap in the literature.

More precisely, we will consider the typical parallel plate setup with PEC and/or PMC boundary conditions.
Extending the construction from \cite{Capri:2012ev} to this configuration, we have to impose extra restrictions on the ansatz for the gauge field.

In the second part of this paper, we will address in explicit detail the boundary gauge invariance in such parallel plate setups.
This boundary gauge invariance occurs when lifting PEC or PMC boundary conditions into the action using auxiliary fields that act as Lagrange multipliers, both in Maxwell and YM theory \cite{Dudal:2020yah,Dudal:2024PEMC,jarnepaper}.
It is also present in the Curci-Ferrari model, but only for PEC plates, giving rise to a discontinuity in the massless limit for PMC \cite{Dudal2026}.
We expect boundary gauge invariance to play an important role when applying the GZ framework to parallel plate setups, since one will need to take into account a boundary horizon condition arising from the boundary gauge symmetry.


\section{Infinitesimal Gribov copies in the presence of PEC or PMC plates: general setup}

We will be considering \(SU(2)\) Yang-Mills theory in Landau gauge \(\p_iA_i^a=0\). For notational ease, we will work in 3D Euclidean space, but the argument can readily be extended to 4D, completely analogous to \cite{Capri:2012ev}.
In 3D, conveniently, both the color indices \(a,b,\dots\) and the space indices \(i,j,\dots\) take values in \(\{1,2,3\}\). We will use the notation \((x_1,x_2,x_3) = (x,y,z)\).
The Yang-Mills action is given by
\begin{equation}
    S_\text{YM} = \frac1{4}\int \dd[3]{x} F_{ij}^a F_{ij}^a,
\end{equation}
and the Hermitian Faddeev-Popov operator by
\begin{equation}\label{FPoperator}
    \M^{ab} = -\left( \delta^{ab}\p^2 - g\varepsilon^{abc} A_i^c\p_i \right),
\end{equation}
where \(F_{ij}^a = \p_i A_j^a - \p_j A_i^a + g \varepsilon^{abc}A_i^b A_j^c\), and \(\varepsilon^{abc}\) is the Levi-Civita symbol, the structure constants for \(SU(2)\).
We will be looking for zero-modes \(\omega\) of this operator, i.e.~
\begin{equation}\label{eq:zero-mode}
    \M^{ab} \omega^b=0.
\end{equation}
Given such a non-trivial zero-mode,  one obtains a pair of infinitesimally related configurations satisfying the Landau condition to first order in the gauge parameter.
Indeed, let \(A\) be a gauge configuration that satisfies the Landau gauge \(\p_iA_i^a=0\), then we can consider a gauge transformation \(\hat A\) of \(A\) defined by
\begin{equation}
    \hat A_i^a = A_i^a - \left( \delta^{ab} \p_i -  g\varepsilon^{abc} A_i^c \right) \omega^b+O(\omega^2).
\end{equation}
Since \(\omega\) is a zero-mode of \(\M\), we see that \(\hat A\) also satisfies Landau gauge, so that, to first order in \(\omega\), it is an infinitesimal Gribov copy of \(A\) :
\begin{equation}
    \p_i \hat A_i^a = \p_i A_i^a + \M^{ab}\omega^b = \p_i A_i^a = 0.
\end{equation}

There are two possible types of normalizability requirement for the gauge field \(A\). The first one is having a finite YM action:
\begin{equation}\label{L2}
    S_\text{YM}[A] < \infty,
\end{equation}
which is a natural condition to demand given the exponential dampening in the Euclidean path integral\footnote{Although this is a delicate point as noted in \cite{Coleman:1978ae}. We thank P.~De Fabritiis for pointing our attention towards this reference.}. The second, stronger, normalizability requirement is that of finite \(L_2\)-norm:
\begin{equation}\label{L2norm}
    ||A||_2^2 = \int \dd[3]{x} A_i^a A_i^a<\infty,
\end{equation}
which is natural to consider in the context of the Gribov region and the fundamental modular region \cite{DellAntonio91}. Reconsidering the Faddeev-Popov operator \eqref{FPoperator}, it is clear that for small enough $gA$, the eigenvalues will be positive as being dominated by the minus Laplacian.  For larger $gA$, a first negative ``bound state'' eigenvalue will appear, and more so on for increasingly strong fields. The (first) Gribov region $\Omega$ is the region of fields such that the eigenvalues are all positive, its boundary, the so-called (first) Gribov horizon $\partial\Omega$, is where the first zero eigenvalue appears.

The boundary configuration we will consider is that of two infinitely long parallel lines (not plates, as we are in 3D) located at \(z=\pm L/2\), and normal vector \(n_i=\delta_{i3}=(0,0,1)\). We will consider lines with perfect magnetic conductor (PMC) boundary conditions that are given by
\begin{equation}\label{pmccond}
    F_{ij}^a n_j = F_{i3}^a = 0 \quad\text{ at }\quad{z=\pm L/2},
\end{equation}
and lines with perfect electric conductor (PEC) boundary conditions
\begin{equation}\label{peccond}
    \tilde F_{i}^a n_i = \frac{\varepsilon_{zij}}2 F_{ij}^a = F_{12}^a = 0 \quad\text{ at }\quad{z=\pm L/2}.
\end{equation}
Note that both PMC and PEC boundary conditions are gauge invariant, as a vanishing field strength is sent to a vanishing field strength under an adjoint gauge transformation.

\section{Infinitesimal Gribov copies in the presence of PEC or PMC plates:  via cylindrical coordinates}
We will base our construction of zero-modes on Henyey's method \cite{Henyey:1978qd}, generalized in \cite{Guimaraes:2011sf,Capri:2012ev,Landim:2012eg}, hereafter worked out in 3D Euclidean space with cylindrical coordinates $(\rho,\varphi,z)$, with the metric
\begin{equation}
\mathrm{d}s^{2}=\mathrm{d}\rho^{2}+\rho^{2}\mathrm{d}\varphi^{2}+\mathrm{d}z^{2}.
\end{equation}
For these constructions in cylindrical coordinates, we will only consider the region between the plates\footnote{If instead one considers $z\in\mathbb{R}$, the $L_2$-norms will acquire a divergent factor $\int\dd{z}$ since the gauge field configurations independent of $z$. In that case, the solution should be interpreted as a configuration with finite quantities per unit length in the $z$-direction.}, given by
\begin{equation}
z\in[-L/2,L/2], \quad \rho\in[0,\infty), \quad \varphi\in[0,2\pi).
\end{equation}

We use generators $\tau_{a}$ of $\mathfrak{su}(2)$ such that
\begin{equation}
[\tau_{a},\tau_{b}]=2\epsilon_{abc}\tau_{c},
\end{equation}
or, equivalently, we work in color components with structure constants $\epsilon^{abc}$.

To avoid metric ambiguities, we write the gauge field in orthonormal cylindrical components:
\begin{equation}
A=A_{\hat{\rho}}\hat{e}_{\rho}+A_{\hat{\varphi}}\hat{e}_{\varphi}+A_{\hat{z}}\hat{e}_{z}.
\end{equation}

\subsection{PMC zero modes using a rotating color basis}

For our first construction of zero-modes, let us define the angle-dependent color basis
\begin{equation}
    \begin{cases}
        T_{1}(\varphi)=\cos(k\varphi)\tau_{1}+\sin(k\varphi)\tau_{2},\\
        T_{2}(\varphi)=-\sin(k\varphi)\tau_{1}+\cos(k\varphi)\tau_{2},\\
        T_{3}=\tau_{3}
    \end{cases}
\end{equation}
with $k\in\mathbb{Z}$.
The condition $k\in\mathbb{Z}$ guarantees the correct monodromy under $\varphi\mapsto\varphi+2\pi$. Moreover,
\begin{equation}
\partial_{\varphi}T_{1}=kT_{2}, \quad \partial_{\varphi}T_{2}=-kT_{1},
\end{equation}
and we have the usual orientation in color space
\begin{equation}
T_{1}\times T_{2}=T_{3}, \quad T_{2}\times T_{3}=T_{1}, \quad T_{3}\times T_{1}=T_{2}.
\end{equation}

Next, consider the gauge field ansatz
\begin{equation}\label{cilan}
A_{\hat{\rho}}=0, \quad A_{\hat{\varphi}}=a(\rho)T_{3}, \quad A_{\hat{z}}=0.
\end{equation}
This configuration is Abelian in color, because the gauge field always points in the $T_{3}$ direction. It is readily checked that this ansatz obeys Landau gauge. Moreover, one can quickly verify that this configuration obeys PMC conditions \eqref{pmccond} everywhere (so in particular on the plates), whereas PEC conditions  are never fulfilled, since \eqref{peccond} is equivalent to $F_{\hat\rho\hat\varphi}^3=0$, which is never the case since it is the only non-vanishing component of the field strength arising from a configuration \eqref{cilan}, see \eqref{pecnonzero}.

We can now construct the zero-modes themselves: take
\begin{equation}
\omega(\rho,\varphi)=\sigma(\rho)T_{1}(\varphi),
\end{equation}
for some function $\sigma(\rho)$.\footnote{Note that since the gauge field ansatz is Abelian in color, the non-triviality of the zero-mode arises from the angular dependence of the zero-mode in the $\tau_{1},\,\tau_2$ directions.}
The zero-mode equation \eqref{eq:zero-mode} then becomes
\begin{equation}
\sigma^{\prime\prime}+\frac{1}{\rho}\sigma^{\prime}-\frac{k^{2}}{\rho^{2}}\sigma-\frac{gk}{\rho}a(\rho)\sigma=0.
\end{equation}
This equation can be solved in the Henyey spirit: choose $\sigma$ and reconstruct $a$,
\begin{equation}\label{vgl-a}
a(\rho)=\frac{\rho}{gk}\left(\frac{\sigma^{\prime\prime}}{\sigma}+\frac{1}{\rho}\frac{\sigma^{\prime}}{\sigma}-\frac{k^{2}}{\rho^{2}}\right).
\end{equation}

Let us make the choice
\begin{equation}
\sigma(\rho)=\frac{\rho^{k}}{1+\rho^{2k}}.
\end{equation}
For $\rho\rightarrow0$, we have $\sigma(\rho)\sim\rho^{k}$, so the zero-mode is regular at the origin for $k\ge1$. For $\rho\rightarrow\infty$, we have $\sigma(\rho)\sim\rho^{-k}$, so it will be normalizable in the transverse plane for $k>1$. Let us compute the corresponding gauge field: working out \eqref{vgl-a}, one gets
\begin{equation}
a(\rho)=-\frac{8k}{g}\frac{\rho^{2k-1}}{(1+\rho^{2k})^{2}}.
\end{equation}
The explicit gauge field configuration is thus
\begin{equation}
A_{\hat{\rho}}=0, \quad A_{\hat{\varphi}}=-\frac{8k}{g}\frac{\rho^{2k-1}}{(1+\rho^{2k})^{2}}T_{3}, \quad A_{\hat{z}}=0,
\end{equation}
and the zero-mode is
\begin{equation}
\omega(\rho,\varphi)=\frac{\rho^{k}}{1+\rho^{2k}}\left(\cos(k\varphi)\tau_{1}+\sin(k\varphi)\tau_{2}\right) .
\end{equation}

The $L^{2}$-norm of the gauge field between the plates is
\begin{equation}
||A||_{2}^{2}=\int_{-L/2}^{L/2}\dd{z}\int_{0}^{2\pi}\dd{\varphi}\int_{0}^{\infty}\dd{\rho}\rho A_{\hat i}^{a}A_{\hat i}^{a}.
\end{equation}
For our ansatz
\begin{equation}
A_{\hat{i}}^{a}A_{\hat{i}}^{a}=a(\rho)^{2},
\end{equation}
up to the normalization of the generators, so it suffices to study
\begin{equation}
\int_{0}^{\infty}\dd{\rho}\rho a(\rho)^{2}.
\end{equation}
For $\rho\rightarrow0$, we have
$\rho a(\rho)^{2}\sim\rho^{4k-1}$
which is integrable at the origin for $k>0$.
For $\rho\rightarrow\infty$, we have
$\rho a(\rho)^{2}\sim\rho^{-4k-1}$,
which is integrable at infinity for $k>0$. Hence
\begin{equation}
||A||_{2}^{2}<\infty \quad \text{for} \quad k\ge1.
\end{equation}

The $L_2$-norm of the zero-mode is
\begin{equation}
||\omega||_{2}^{2}=\int_{-L/2}^{L/2}\dd{z}\int_{0}^{2\pi}\dd{\varphi}\int_{0}^{\infty}\dd{\rho}\rho\sigma(\rho)^{2},
\end{equation}
for which one can check analogously that
\begin{equation}
||\omega||_{2}^{2}<\infty \quad \text{for} \quad k\ge2.
\end{equation}

Since the gauge field always points in the color direction $T_{3}$, the non-Abelian term $g\epsilon^{abc}A_{i}^{b}A_{j}^{c}$ in the field strength vanishes. The only non-zero field-strength component is
\begin{equation}\label{pecnonzero}
F_{\hat{\rho}\hat{\varphi}}^{3} =\frac{1}{\rho}\partial_\rho( \rho a(\rho))=-\frac{16k^{2}}{g}\frac{\rho^{2k-2}(1-\rho^{2k})}{(1+\rho^{2k})^{3}}.
\end{equation}
Therefore the three-dimensional Yang-Mills action goes like
\begin{equation}
S_{YM}\propto\int_{-L/2}^{L/2}\dd{z}\int_{0}^{2\pi}\dd{\varphi}\int_{0}^{\infty}\dd{\rho}\rho(F_{\hat{\rho}\hat{\varphi}}^{3})^{2},
\end{equation}
such that one can check that
\begin{equation}
S_{YM}<\infty \quad \text{for} \quad k\ge1.
\end{equation}
In conclusion, if one chooses $k\ge2$, one simultaneously obtains finite action, finite $L_{2}$-norm of the gauge field, and normalizability of the zero-mode.

It remains to study whether this family actually belongs to the Gribov region in the strong sense, namely how the full spectrum of the FP operator behaves around this configuration. The presence of a zero-mode shows that the configuration lies on the Gribov horizon, but it does not by itself determine the positivity of all other modes.

\subsection{PMC zero modes in the background of a thick center vortex}
Another interesting class of (topologically non-trivial) copies were discussed in \cite{Maas:2005qt} for center vortices, in cylindrical coordinates given by the only non-zero component
\begin{eqnarray}\label{vor1}
A_\varphi^a=\frac{1}{g}\delta^{a3}\frac{\mu(\rho)}{\rho},
\end{eqnarray}
where the profile $\mu(\rho)$ is supposed to be sufficiently smooth, with $\mu(0)=0$ and $\mu(\infty)=2n+1$, $n\in\mathbb N$, related to the flux. Let us refer to \cite{Diakonov:1999gg,Diakonov:2002bx} for more details.  As the configuration \eqref{vor1} implies the following field strength tensor components,
\begin{eqnarray}\label{vor2}
F_{12}^a= \frac{1}{g}\frac{\partial_\rho \mu(\rho)}{\rho}, \quad F_{23}^a=F_{13}^a=0,
\end{eqnarray}
we immediately see that the PMC boundary condition \eqref{pmcb} is trivially fulfilled, whilst the PEC one \eqref{pecb} can never be.

We conclude that the center vortex results of \cite{Maas:2005qt} immediately transfer to the Yang-Mills case with a center vortex combined with parallel PMC plates.

Although cylindrical coordinates are quite natural to consider in the context of parallel wires, we were not able so far to construct gauge field configurations obeying the PEC boundary condition. In the next section, we will remedy this situation by using a construction in spherical coordinates.

\section{Infinitesimal Gribov copies in the presence of PEC or PMC plates: via spherical coordinates}


\subsection{Analytical preparation}

The beginning of our discussion closely follows the construction in three dimensions in \cite{Capri:2012ev}. Since we do not aim to find all zero-modes, we can make some ansatzes that simplify our calculations.
Thus, instead of working with a completely general gauge configuration, we will start from a well-chosen type of gauge field:
\begin{equation}\label{eq:ansatz-A}
    A_i^a = \varepsilon_{aij} x_j h(r).
\end{equation}
Here, we introduced spherical coordinates \((r,\varphi,\theta)\) for Euclidean space, and \(h\) is an arbitrary function of the radial distance.
Note that color and space indices are being mixed in the Levi-Civita symbol, which is not a problem since both take values in \(\{1,2,3\}\).
For such gauge configurations, the differential equations \eqref{eq:zero-mode} for the zero-modes become
\begin{equation}\label{eq:zero-mode-PDE}
    \p^2 \omega^a + g h(r) \varepsilon^{abc} \varepsilon_{cij} x_i  \p_j\omega^b = 0,
\end{equation}
in which we recognize the angular momentum operator
\begin{equation}
    L_c = \varepsilon_{cij} x_i  \p_j.
\end{equation}
We now make a second ansatz: we will only consider zero-modes \(\omega\) of the form
\begin{equation}
    \omega^a = (\omega^{(1)}, \omega^{(2)},0),
\end{equation}
such that the differential equations \eqref{eq:zero-mode-PDE} reduce to
\begin{equation}
    \begin{cases}
        \p^2 \omega^{(1)} +g h(r) L_3 \omega^{(2)} =0  \\
        \p^2 \omega^{(2)} -g h(r) L_3 \omega^{(1)} =0.
    \end{cases}
\end{equation}
Because \(L_3 = \p/\p\varphi\), we immediately find a class of solutions of the form
\begin{equation}
    \begin{cases}
        \omega^{(1)}(r,\varphi,\theta) = \sigma(r) \sin\varphi\sin\theta \\
        \omega^{(2)}(r,\varphi,\theta) = \sigma(r) \cos\varphi\sin\theta,
    \end{cases}
\end{equation}
where \(\sigma(r)\) needs to obey the single ODE
\begin{equation}
    \sigma''(r) +\frac2r \sigma'(r)-\frac2{r^2}\sigma(r) = g h(r) \sigma(r).
\end{equation}
Performing the substitution \(\psi(r) = r\sigma(r)\), this ODE is recast as
\begin{equation}\label{eq:ODE-empty-space}
    h(r) = \frac{1}{g\psi(r)}\left( \psi''(r) + \frac4r \psi'(r) \right).
\end{equation}
In empty space, the function \(h(r)\) is still completely free, cfr.~\cite{Capri:2012ev}.
But if there is a PMC plate present at \(z=\pm L/2\), we get some extra restrictions on \(h(r)\) there:
\begin{align}\label{pmcb}
    F_{i3}^a     &=\p_i A_3^a - \p_3A_i^a + g \varepsilon^{abc} A_i^b A_3^c \nonumber\\
    &=\p_i(\varepsilon_{a3j} x_j h(r)) - \p_3 (\varepsilon_{aij} x_j h(r)) + g \varepsilon^{abc} (\varepsilon_{bij} x_j h(r))(\varepsilon_{c3k} x_k h(r)) \nonumber\\
    &=-2\varepsilon_{ai3}h(r) + \frac{h'(r)}{r}x_j (\varepsilon_{a3j} x_i - \varepsilon_{aij} z) + g \varepsilon_{i3k} x_a x_k h(r)^2=0&\text{ at }{z=\pm L/2}.
\end{align}
For \(i=3\), this equation is automatically satisfied, but for \(i=1\) we get two distinct equations
\begin{equation}\label{restricties}
    \begin{cases}
        \frac{h'(r)}{r} = -g h(r)^2 \\
        2h + \frac{h'(r)}{r}(x^2+z^2)-gy^2h^2 =0
    \end{cases}
    \hfill\text{ at }{z=\pm L/2}.
\end{equation}
Filling in the first equation in the second, we find the restriction
\begin{equation}\label{eq:restriction-h}
    h(r) = \frac{2}{g}\frac{1}{r^2} \quad\text{ at }{z=\pm L/2},
\end{equation}
which solves the first equation as well.
For \(i=2\), we get a similar set of equations with the exact same restriction on \(h(r)\).

Similarly, if there is a PEC plate at \(z=\pm L/2\), we have the restrictions
\begin{align}\label{pecb}
    F_{12}^a  &=\p_1 A_2^a - \p_2 A_1^a + g \varepsilon^{abc} A_1^b A_2^c\\
    &=\p_1(\varepsilon_{a2j} x_j h(r)) - \p_2 (\varepsilon_{a1j} x_j h(r)) + g \varepsilon^{abc} (\varepsilon_{b1j} x_j h(r))(\varepsilon_{c2k} x_k h(r)) \nonumber \\
    &=-2\delta_{a3}h(r) + \frac{h'(r)}{r}x_j (\varepsilon_{a2j} x - \varepsilon_{a1j} y) + g x_a z h(r)^2=0&\text{ at }{z=\pm L/2}.
\end{align}
The equation for \(a=1\) and \(a=2\) both yields
\begin{equation}
    \frac{h'(r)}{r} = -g h(r)^2,
\end{equation}
whereas the equation for \(a=3\) reads
\begin{equation}
    2h + \frac{h'(r)}{r}(x^2+y^2)-gz^2h^2 =0.
\end{equation}
Again, filling in the first equation in the second, we find the exact same restriction \eqref{eq:restriction-h} as for PMC plates.

Since this restriction only holds at \(z=\pm L/2\), we still have some freedom left for \(r<L/2\).
Indeed,
\begin{equation}\label{eersteh}
    h(r) \coloneq \frac2g \frac1{r^2} + \tilde h(r) \mathbbm{1}_{r<L/2}(r)
\end{equation}
also satisfies the restriction \eqref{eq:restriction-h}.\footnote{Note that adding a term $\propto\mathbbm{1}_{r>L/2}(r)$ is not possible. Indeed, this follows from the simple geometric observation that any sphere with radius larger than $L/2$ will intersect the boundary lines at \(z=\pm L/2\) and hence mess up the constraint \eqref{eq:restriction-h}.}
The function \(\tilde h(r)\) contains some freedom, but there are restrictions on it as well.
First, since we want \(h\) to be a smooth function, we have to demand that
\begin{equation}
    \tilde h(r) \xrightarrow[\text{smoothly}]{r\nearrow L/2} 0.
\end{equation}
In principle, for our purposes, it suffices for \(h(r)\) to be continuously differentiable, but our method allows one to demand higher degrees of smoothness if required.
Second, because \(A\) is proportional to \(h(r)\), see Eq.~\eqref{eq:ansatz-A}, we also do not want \(h\) to have a singularity in the origin. Therefore, we have to demand that
\begin{equation}
    \tilde h(r) \xrightarrow{r\searrow 0} -\frac2g\frac{1}{r^2}.
\end{equation}
Both demands are satisfied by any function \(\tilde h(r)\) of the form
\begin{equation}\label{free1}
    \tilde h(r) = -\frac2g\frac{1}{r^2}\left(\frac{L/2-r}{L/2}\right)^n \times \left(1+\sum_{i=1}^{m}c_ir^i \right),
\end{equation}
where the larger \(n\in\mathbb{N}\) is, the smoother \(h(r)\) is at \(r = L/2\). The rightmost factor has been included in order to give us some free parameters \(c_i\) that we will need later on. The larger we choose \(m\in\mathbb{N}\), the more degrees of freedom we have available.

Having worked through the restrictions that arise because of the plates, we find that the ODE \eqref{eq:ODE-empty-space} we need to solve has become
\begin{equation}\label{eq:ODE-with-plates}
    \frac2{r^2} \left[ 1 - \left(\frac{L/2-r}{L/2}\right)^n \left(1+\sum_{i=1}^{m}c_ir^i \right) \mathbbm{1}_{r<L/2}(r) \right] = \frac{1}{\psi(r)}\left( \psi''(r) + \frac4r \psi'(r) \right).
\end{equation}
Unfortunately, this ODE does not have a solution in closed form for any relevant values of \(n,m\). We will therefore resort to numerical methods for finding solutions.


\subsection{Numerically solving the zero-mode ODE}

We will divide space in three separate regions. First, `region 0' close to the origin, \(r<\varepsilon\) for some small \(\varepsilon\), where \(\psi_0(r)\) solves the ODE \eqref{eq:ODE-with-plates} which reads
\begin{equation}\label{eq:ODE-region0}
    \frac2{r^2} \left[ 1 - \left(\frac{L/2-r}{L/2}\right)^n \left(1+\sum_{i=1}^{m}c_ir^i \right) \right] = \frac{1}{\psi_0(r)}\left( \psi_0''(r) + \frac4r \psi_0'(r) \right).
\end{equation}
Second, `region 1' of a pierced sphere in between the plates, \(\varepsilon < r <L/2\), with solution \(\psi_1(r)\) of the same ODE
\begin{equation}\label{eq:ODE-region1}
    \frac2{r^2} \left[ 1 - \left(\frac{L/2-r}{L/2}\right)^n \left(1+\sum_{i=1}^{m}c_ir^i \right) \right] = \frac{1}{\psi_1(r)}\left( \psi_1''(r) + \frac4r \psi_1'(r) \right).
\end{equation}
Lastly, the outer `region 2' given by \(r>L/2\), where \(\psi_2(r)\) solves the simpler ODE
\begin{equation}\label{eq:ODE-region2}
    \frac2{r^2} = \frac{1}{\psi_2(r)}\left( \psi_2''(r) + \frac4r \psi_2'(r) \right).
\end{equation}

In region 0, we can expand \(\psi_0(r)\) as a series around \(r=0\):
\begin{equation}\label{eq:psi0-series}
    \psi_0(r) = 1 + \sum_{i=1}^k a_i r^i,
\end{equation}
where the number of terms $k$ can be chosen depending on how closely one wishes the series to approach the exact solution \(\psi_0\) for \(r\approx0\).
We have set the constant term equal to 1, since this corresponds to the normalization of \(\psi_0\) (which is free to choose because of the homogeneity of the ODE).
Filling in the series \eqref{eq:psi0-series} in the ODE \eqref{eq:ODE-region0}, we can fix \(a_i\) in terms of the \(c_j\) such that \(\psi_0(r)\) fulfills the ODE in region 0, i.e.~\(r<\varepsilon\) (with accuracy up to an order \(\mathcal{O}(r^{k-1})\) of choice).

Now, let us extend the solution \(\psi_0\) further away from the origin, to region 1: \(\varepsilon<r<L/2\). The solution \(\psi_1(r)\) has to be glued to \(\psi_0(r)\) at \(r=\varepsilon\). For this illustration, we match the solutions and their first derivative (but higher order smoothness can be demanded if required):
\begin{equation}
    \begin{cases}
        \psi_1(\varepsilon) = \psi_0(\varepsilon) \\
        \psi_1'(\varepsilon) = \psi_0'(\varepsilon).
    \end{cases}
\end{equation}
With these boundary conditions the ODE \eqref{eq:ODE-region1} in region 1 can readily be solved numerically, such that we have obtained the solution \(\psi_1(r)\) for \(\varepsilon<r<L/2\).

Finally, we consider region 2 with its ODE \eqref{eq:ODE-region2}. The general solution for this ODE is \(b_1 r^{-\frac32 - \frac{\sqrt{17}}{2}} + b_2 r^{-\frac32 + \frac{\sqrt{17}}{2}}\), but since we need a normalizable solution, we set
\begin{eqnarray}
    \psi_2(r) = b_1 r^{-\frac32 - \frac{\sqrt{17}}{2}}.
\end{eqnarray}
One can now glue \(\psi_1\) and \(\psi_2\) at \(r=L/2\) using the free coefficients \(b_1,c_1,\ldots,c_m\).

\begin{figure} \centering
\begin{subfigure}[t]{0.3\textwidth}
\centering \includegraphics[width=\linewidth]{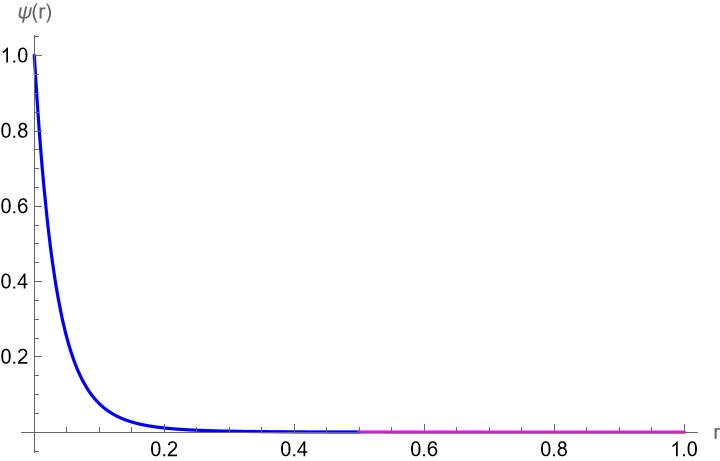}
\caption{} \end{subfigure} \hfill
\begin{subfigure}[t]{0.3\textwidth} \centering
\includegraphics[width=\linewidth]{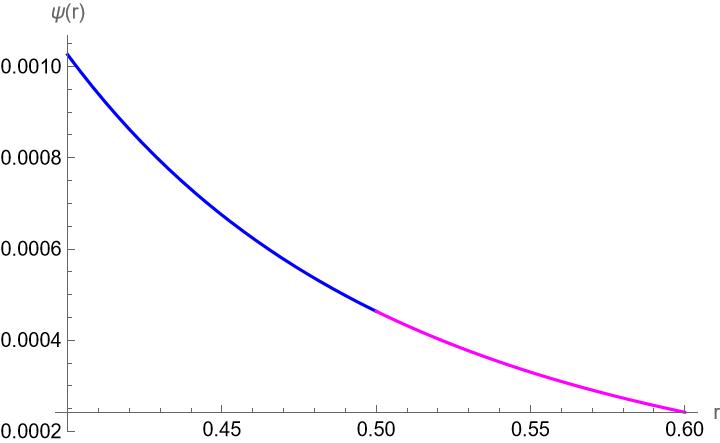} \caption{}
\end{subfigure} \hfill \begin{subfigure}[t]{0.3\textwidth}
\centering \includegraphics[width=\linewidth]{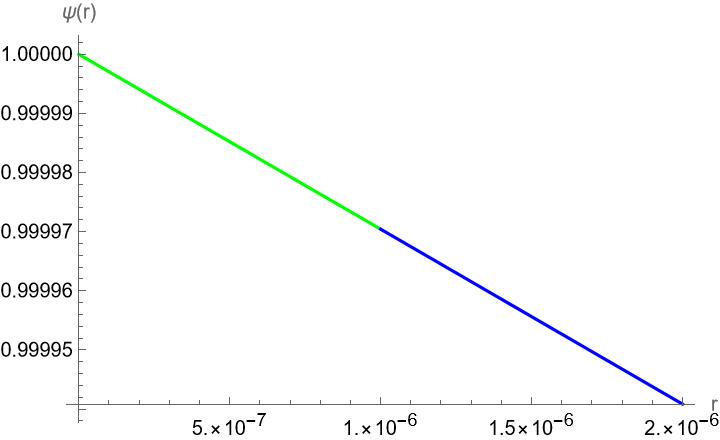} \caption{}
\end{subfigure} \caption{Plot of the solution using $h(r)$ as in \eqref{eersteh}, for $L=1$, $\varepsilon=10^{-6}$. $\psi_0(r)$ is shown in green, $\psi_1(r)$ in blue, $\psi_2(r)$ in magenta. (a) full view of the domain $0<r<L$, (b) zoom in around $r=L/2$, (c) zoom in around $r=\varepsilon$.} \label{fig:plots1} \end{figure}

For the numerical evaluation, we set \(\varepsilon=10^{-6}\) and the number of terms \(n=5,m=2,k=3\), as this choice will yield a more than sufficiently smooth function \(h(r)\) while also having enough degrees of freedom \(c_i\). We plot the illustrative solution for $L=1$ in Fig.~\ref{fig:plots1}.

Let us now inspect the normalizability properties of this solution.
To start, note that for large $r$ we have \(\psi(r) \sim r^{-\frac32 - \frac{\sqrt{17}}{2}} \approx r^{-3.56}\), such that \(\sigma(r) = \psi(r)/r \sim r^{-\frac52 - \frac{\sqrt{17}}{2}} \approx r^{-4.56}\) and we have zero-modes with finite $L_2$-norm:
\begin{equation}
    ||\omega^a||_2^2 \approx \int_0^\infty \dd{r} r^2 (r^{-4.56})^2 <\infty.
\end{equation}
Next, consider $h(r)$. By construction, it has no singularity in the origin, so we only need to inspect its behavior for \(r\rightarrow \infty\). Looking at Eq.~\eqref{eq:ODE-empty-space}, we have that \(h(r) \sim 1/r^2\) for large $r$. Therefore, our gauge configuration \(A_i^a = \varepsilon_{aij} x_j h(r)\sim 1/r\) for large $r$, and the field strength \(F_{ij}^a \sim 1/r^2\).
We thus have finite YM action
\begin{equation}
    \int \dd[3]{x} F_{ij}^a F_{ij}^a \sim \int_0^\infty \dd{r} r^2 (r^{-2})^2 < \infty,
\end{equation}
but not finite $L_2$-norm for the gauge field:
\begin{equation}
    \int \dd[3]{x} A_i^a A_i^a \sim \int_0^\infty \dd{r} r^2 (r^{-1})^2 = \infty.
\end{equation}
Note that having finite $L_2$-norm is \emph{not} an essential prerequisite to have a meaningful path integral.
The finite $L_2$-norm restriction comes into play when considering the Gribov region $\Omega$,
\begin{equation}
    \Omega=\left\{A_\mu^a\left|\p_\mu A_\mu^a=0,~\M^{ab}>0\right.\right\},
\end{equation}
which can be redefined as the region of (local) minima of the $L_2$-norm \eqref{L2norm} of a generic gauge field when varying it under gauge transformations. The latter definition is the one that enters the proof of the fact that every gauge orbit crosses $\Omega$, see \cite{DellAntonio91}. Of course, that  proof is in the absence of boundaries.

The infinitesimal copies identified through these normalizable zero modes are therefore relevant to the local Gribov problem, although they do not belong to the Gribov region in the strict sense. It remains an open question if every gauge orbit still passes through $\Omega$ in the presence of planar boundaries.
 
Here is also a good place to remind the reader  that we are exclusively focusing on infinitesimal Gribov copies, compatible with either PEC or PMC boundary conditions. This is not without reason: the concept of the Gribov region $\Omega$ and the restriction to it are based on the zero modes of the Faddeev-Popov operator. 
This, in turn, originates from Gribov’s observation (and proving) that for any gauge field ``close enough'' to the first Gribov horizon, there is a copy on the other side, see \cite{Gribov78,Sobreiro05}. ``Close enough'' evidently corresponds to field configurations connected through an infinitesimal (normalizable) gauge transformation. 

At the formal level, the local near-horizon argument can be carried over to the present setting because the PEC and PMC conditions are gauge covariant. This statement assumes that the gauge parameters obey the same regularity and normalizability requirements used in defining the relevant Faddeev-Popov operator \cite{Gribov78,Sobreiro05}. As such, it remains a natural idea to restrict the gauge field integration domain to the inside of the first Gribov horizon, irrespective of PEC, PMC or no boundary conditions.

Searching for finite copies requires a different analysis and is currently under study, requiring a non-trivial generalization of e.g.~the Gribov pendulum \cite{Gribov78} or the Henyey classs of finite copies \cite{Henyey:1978qd}, see also \cite{Sobreiro05} to the case with PEC or PMC conditions on top.  We stress however that, the aforementioned finite copies satisfie neither PEC nor PMC boundary conditions.

\subsection{Infinitesimal Gribov copies and zero-modes with finite $L_2$-norm}
Reconsidering the constraints \eqref{restricties}, which validate both PEC and PMC boundary conditions, there is actually also the trivial solution where $h(r)=0$ in a region around $z=\pm L/2$, such that its derivative vanishes at $z=\pm L/2$ as well. Therefore, consider e.g.~the following very simple ansatz,
\begin{equation}\label{gecutteh}
h(r)=\left\{
  \begin{array}{ll}
    \left(\frac{L/4-r}{L/4}\right)^n \left(1+\sum_{i=1}^{m}c_ir^i \right) & \text{for}~r\leq L/4 \\
    0 & \text{for}~ r>L/4
  \end{array},
\right.
\end{equation}
where $n\in \mathbb{N}_0$ can be chosen at will to promote the smooth decay to zero for $r\to L/4$, and the number of free coefficients $m$ can be chosen at will to achieve the desired degree of smoothness. The boundary conditions \eqref{restricties} at $z=\pm L/2$ are trivially fulfilled. The profile $h(r)$ is evidently regular near the origin.

To find the corresponding gauge field configuration, we must solve \eqref{eq:ODE-empty-space}, which for the `outer region' $r>L/4$ collapses to
\begin{equation}\label{eq:ODE-region2bis}
     \psi_2''(r) + \frac4r \psi_2'(r)=0.
\end{equation}
A normalizable solution reads
\begin{equation}\label{opl1}
    \psi_2(r)= \frac{b}{r^3}
\end{equation}
with $b$ an integration constant that we will fix later by matching the two parts of the solution. Indeed, in the `inner region' $r<L/4$, we must solve
\begin{equation}\label{eq:ODE-region0bis}
          \left(\frac{L/4-r}{L/4}\right)^n \left(1+\sum_{i=1}^{m}c_ir^i \right) = \frac{1}{\psi(r)}\left( \psi''(r) + \frac4r \psi'(r) \right).
\end{equation}
As before, we can first solve this for $r<\varepsilon$ using a series as in \eqref{eq:psi0-series}  (`region 0') and then glue again smoothly to the numerical solution for larger $r$ in `region 1'.

\begin{figure}
    \centering

    \begin{subfigure}[t]{0.3\textwidth}
        \centering
        \includegraphics[width=\linewidth]{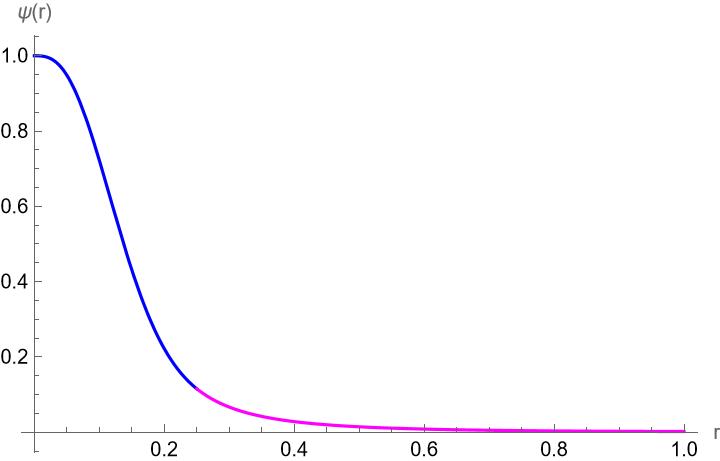}
        \caption{}
    \end{subfigure}
    \hfill
    \begin{subfigure}[t]{0.3\textwidth}
        \centering
        \includegraphics[width=\linewidth]{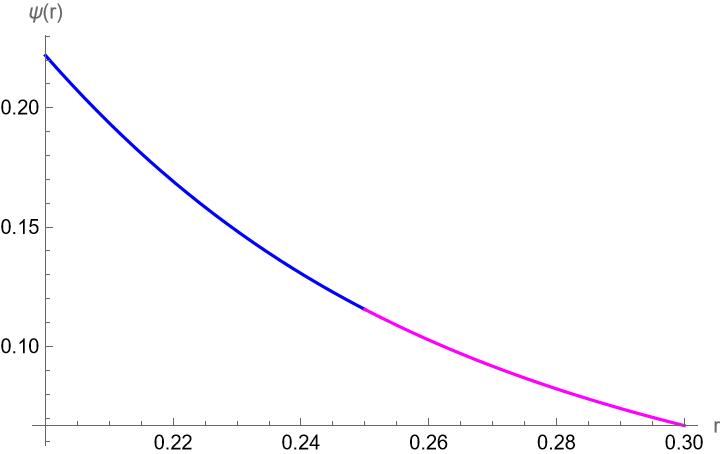}
        \caption{}
    \end{subfigure}
    \hfill
    \begin{subfigure}[t]{0.3\textwidth}
        \centering
        \includegraphics[width=\linewidth]{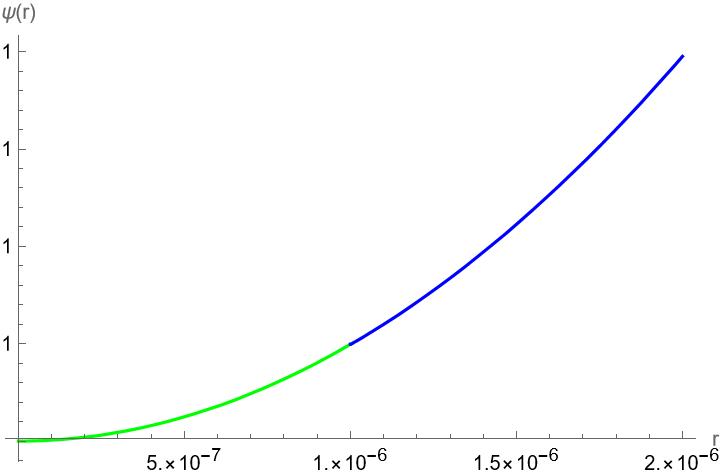}
        \caption{}
    \end{subfigure}

    \caption{Plot of the solution using $h(r)$ as in \eqref{gecutteh}, for $L=1$, $\varepsilon=10^{-6}$. $\psi_0(r)$ is shown in green, $\psi_1(r)$ in blue, $\psi_2(r)$ in magenta. (a) full view of the domain $0<r<L$, (b) zoom in around $r=L/4$, (c) zoom in around $r=\varepsilon$.}
    \label{fig:plots2}
\end{figure}

Let us present a concrete example. We take $n=2$ and $m=1$, then
\[\psi_0(r)=1+\frac{r^2}{10}+(c_1-8)\frac{r^3}{18}+\mathcal{O}(r^4)\]
obeys \eqref{eq:ODE-region0bis} for $r<\varepsilon$ up to $O(r^2)$.
We can then numerically find a solution \(\psi_1\) of \eqref{eq:ODE-region0bis} in $\varepsilon<r<L/4$ that is glued to \(\psi_0\) in \(r=\varepsilon\) in a continuously differentiable manner. Finally, one has to fix the values of $b$ and $c_1$ such that the numerical solution $\psi_1$ is glued to $\psi_2$ in $r=L/4$, again continuously differentiable. We plot the solution resulting from this procedure in Fig.~\ref{fig:plots2}.

From the above results, it is immediate that the zero-mode $\omega^a(r)$ is both sufficiently regular around the origin and sufficiently fast decaying for large $r$, namely $\omega(r)\sim r^{-4}$, to guarantee a finite $L_2$-norm. Moreover, from the regularity at the origin of $h(r)$ defined by \eqref{gecutteh}  and its vanishing for $r>L/4$, the corresponding gauge field configuration will exhibit both finite $L_2$-norm and action.

Needless to say, one can imagine infinitely many configurations of the type \eqref{gecutteh} to construct well-behaved zero-modes and gauge copies, obeying both PMC and PEC conditions at $z=\pm L/2$. Geometrically the picture is clear: the original gauge field is (strictly) localized in between the plates, so that the homogeneous PEC or PMC boundary conditions are per se trivially satisfied.

For completeness, let us point out that we implemented a ``sufficient degree of smoothness'' for (numerical) simplicity by means of the parameters $n$ and $m$ in \eqref{free1} or \eqref{gecutteh}, but by using bump functions with appropriate supports, in principle $C^\infty$-solutions are also achievable.


\section{Emergent boundary gauge invariance in Yang-Mills theory}

Let us now turn to the study of boundary gauge invariance in four-dimensional (Euclidean) YM theory, \(d=4\). In \cite{Dudal:2020yah,Dudal:2024PEMC,jarnepaper}, this additional boundary gauge invariance was encountered already in Maxwell theory, where it was noted that the boundary multiplier fields \(b\) exhibit a gauge symmetry themselves, see also \cite{Karabali:2025olx}. More concretely, after having integrated out the gauge field, the resulting effective boundary action in $d-1$ dimensions was seen to be invariant under boundary gauge transformations \(b_i \rightarrow b_i + \p_i \omega\), for \(\omega(\cev x)\) an arbitrary function living on the boundary.
Fixing this boundary gauge freedom in effect kills one degree of freedom, such that of the \(d-1\) auxiliary \(b\)-field polarizations, only \(d-2\) contribute to the Casimir energy.

In \cite{Dudal2026}, the same phenomenon was found in the Abelian approximation of the Curci-Ferrari (CF) model \cite{CurciFerrari76} in Landau gauge, albeit only for PEC plates, not for PMC plates. We remind here that the Curci-Ferrari model is frequently used to model non-perturbative (infrared) physics of Yang-Mills theory in the Landau gauge by adding by hand a gluon mass term to the gauge-fixed action \cite{Pelaez21}.

Recall that in Maxwell theory, PEC and PMC plates yield the same Casimir energy, and that in the massless free limit, Curci-Ferrari reduces to (\(N_c^2-1\) copies of) Maxwell theory. Therefore, PEC and PMC having different Casimir energy in Curci-Ferrari but not in Maxwell, gives rise to a van Dam--Veltman--Zakharov-like discontinuity \cite{vanDam:1970vg,Zakharov:1970cc} for PMC plates in the CF case.

In this section we want to discuss this boundary gauge invariance in more depth for genuine Yang-Mills theory, explicitly going beyond the Maxwell (Abelian) limit. Since we explicitly showed the existence of normalizable zero modes of the Faddeev–Popov operatorin the presence of PMC and/or PEC plates, one must address the gauge fixing ambiguity, as usual. One motivated way to do so is via the Gribov-Zwanziger (GZ) framework \cite{Gribov78,Zwanziger89,Vandersickel:2012tz}. With the GZ procedure, a good understanding of boundary gauge symmetries is indispensable, since both the bulk and emergent boundary local symmetry will have to be taken into proper account.
We aim to study this observation and its downstream effects in future work.


\subsection{PEC plates}
We consider the usual setup of two parallel, zero thickness plates \(\Sigma_\pm\) at $z=\pm L/2$, now in \(d=4\) dimensions. We first consider PEC boundary conditions on the plates:
\begin{equation}
    \widetilde F_{\mu \nu}^a n_\nu \bigg\vert_{\Sigma_\pm} = 0, \label{pecvgl}
\end{equation}
with corresponding (Euclidean) action
\begin{eqnarray}\label{action}
S=\frac{1}{4}\int \dd[d]{x} F_{\mu\nu}^aF_{\mu\nu}^a + \int \dd[d]{x} \left[ b_i^{a,\pm}(\cev x) n_\nu \widetilde F_{i\nu}^a\delta_\pm\right]
\end{eqnarray}
where we used the shorthand $\delta_\pm=\delta(z\mp \frac L 2)$. $\cev x$ denotes the $d-1$ directions orthogonal to the $z$-direction, with ${\rm d}^dx = {\rm d}^{d-1}\cev x \, {\rm d}z$. In practice, $n_\mu=(\cev 0,1)$, next to $\widetilde F_{\mu\nu}^a= \frac{i}{2} \varepsilon_{\mu\nu\alpha\beta} F_{\alpha\beta}^a$ in Euclidean convention.   We will use indices $i,j,\ldots$ to refer to the $\cev x$-coordinates.

Clearly, \eqref{action} enjoys the standard local gauge invariance, naturally extended to the auxiliary field sector via the adjoint representation
\begin{eqnarray}\label{g1}
\delta A_\mu^a = D_\mu^{ab}[A] \xi^b,\qquad \delta b_i^{a,\pm}= gf^{abc} b_i^{b,\pm} \xi^c.
\end{eqnarray}
However, \eqref{action} obeys another independent local symmetry ``living'' on the plates, generated by
\begin{eqnarray}\label{g2}
\Delta b_i^{a,\pm}(\cev x) = D_i^{ab}[A] \omega^{b,\pm}(\cev x),\qquad \Delta A_\mu^a=0
\end{eqnarray}
which is immediate to verify via the Bianchi identity. Note that the $A$ appearing in the above covariant derivative will automatically be restricted to $z=\pm L/2$ thanks to the $\delta_\pm$ appearing in the action.

When gauge fixing, both local symmetries thus need to be gauge fixed independently. At leading order (Abelian approximation), we already encountered this symmetry in practice in previous work \cite{Dudal2026}, see also \cite{Karabali:2025olx}. Here, we thus see that it extends to the non-Abelian sector. Note that the explicit form of \eqref{g2} at the effective action level of the $b^\pm$-fields---that is, after integrating out the $A$-field---might get quite ugly. Fortunately, we will never need that explicit on-shell version.

Thinking about the Gribov problem and its Gribov-Zwanziger resolution, it means we will have a GZ analysis in the $d$-dimensional bulk related to \eqref{g1} and one in the $(d-1)$-dimensional boundary related to \eqref{g2}. Both will eventually enter the Casimir energy/force. We plan to study the implications in GZ in future work.


\subsection{PMC plates}
In the case of PMC boundary conditions on the plates,
\begin{align}
    F_{\mu \nu}^a n_\nu \bigg\vert_{\Sigma_\pm} = 0, \label{pmcvgl}
\end{align}
and thus
\begin{eqnarray}\label{action2}
S=\frac{1}{4}\int \dd[d]{x} F_{\mu\nu}^aF_{\mu\nu}^a + \int \dd[d]{x} \left[ b_i^{a,\pm}(\cev x) n_\nu F_{i\nu}^a\delta_\pm\right],
\end{eqnarray}
things get more complicated as we no longer have the Bianchi identity at our disposal.

Nonetheless, similar to \eqref{g2}, there is also an independent local boundary symmetry, now generated by
\begin{eqnarray}\label{g2bis}
\Delta b_i^{a,\pm}(\cev x) = D_i^{ab}[A] \omega^{b,\pm}(\cev x),\qquad \Delta A_\mu^a= \theta_\mu^{a,\pm}(\cev x)\delta_\pm.
\end{eqnarray}
We will momentarily check this, upon proper identification of the yet unknown $\theta_\mu^{a,\pm}$. If one prefers to work with smooth, rapidly decaying field configurations rather than those containing a $\delta_\pm$, the latter can be replaced by e.g.~their Gaussian resolution with arbitrary small widths without affecting the final conclusion for vanishing width.

Anyhow, varying the action \eqref{action2} under \eqref{g2bis} leads to
\begin{eqnarray}\label{DeltaS}
  \Delta S = -\int \dd[d]{x} \left[D_i^{ab}F_{i\nu}^b \theta_\nu^{b,\pm}\delta_\pm+\omega^{a,\pm} n_\nu D_i^{ab} F_{i\nu}^b\delta_\pm \right]
  + \int \dd[d]{x} \left[ b_i^{a,\pm}  n_\nu \frac{\delta F_{i\nu}^a}{\delta A_\rho^b}  \times (\theta_\rho^{b,+}\delta_+ +\theta_\rho^{b,-}\delta_-)\delta_\pm \right]
\end{eqnarray}
where here and later on summation over $\pm$ is understood.  We note that the variation of the boundary condition terms is proportional to $D_i^{ab} F_{i\nu}^b$, which are essentially the equations of motion of the $A_\mu^a$-field. This is evidently reflected in the choice for $\Delta A_\mu^a$ in \eqref{g2bis}.


\subsubsection{In dimensional regularization}
Since we are dealing with a gauge theory, dimensional regularization is the natural choice to respect gauge invariance. Recently, in \cite{Vercauteren:2026kdz} some objections were raised against the use of dimensional regularization in the case of infinitely thin boundaries, that is, e.g.~conditions imposed at $z=\pm \frac L 2$, since the $z$-coordinate refers to exactly 1 dimension, not $1-\varepsilon$. Despite the fact that infinitely thin plates are a mathematically idealized concept, these are standard to study the Casimir effect, with or without quantum field theory tools. In principle, we may rely on $\delta^{(n)}(0)=0$ for any $n>0$, see e.g.~\cite[Eq. (4.2.6)]{Collins:1984xc} or \cite[below Eq.~(10.9)]{Zinn-Justin:2002ecy}. Mark that this $n$ does not need to be $n-\varepsilon$ to draw this conclusion ($n=1$ for us), this holds by the very definition and properties of dimensional regularization, in \emph{any} dimension different from $0$.  As an alternative, one might consider working in exactly $d=4$ dimensions\footnote{Note that this also avoids the delicate issue of having to deal with the Levi-Civita tensor in a dimensionality not strictly~$4$.} and ``split'' the overall dimension in $1-\varepsilon$ for the plates (with $z_\mu=\pm \frac L 2$ for $\mu\in[3+\varepsilon, 4]$), and $3+\varepsilon$ for the remaining orthogonal directions. We verified that this keeps everything in fact explicitly finite in leading order Casimir effect computations based on the boundary field technology.

As such, the last line in \eqref{DeltaS} vanishes, as for $L>0$, anyhow $\delta_+\delta_-=0$, while $\displaystyle\int \dd[d]{x} (\ldots)\delta_+\delta_+ \propto \delta(0) = 0$. This means that by adopting
\begin{equation}\label{DeltaS2}
  \theta_\mu^{a,\pm}(\cev x)= -n_\mu \omega^{a,\pm}(\cev x)
\end{equation}
we do get $\Delta S=0$.

Note that also other types of regularization can give $\delta(0)=0$, combine for example the 4th and 5th integral of \cite[eq.(A6)]{Dudal:2020yah}.


\subsubsection{A more careful look}
In case one feels uncomfortable to use regularizations that set $\delta(0)=0$, let us also provide an alternative reasoning taking into explicit account the appropriate counterterm(s), keeping track of the $\delta(0)$-divergence. One can for example introduce a UV cut-off $\Lambda$ as intermediate regulator, for which $\delta(0)\propto\Lambda$.

We remind here about the Lowenstein procedure\footnote{Which in itself is part of BPHZ(L) regularization/renormalization approach, but similar comments apply in case of cut-off regularization.} \cite{Velo:1976gh}: introducing $\Lambda$ in a gauge theory will inevitably lead to e.g.~mass divergences $\propto \Lambda^2$ which seem to violate gauge invariance, or better said, the Slavnov-Taylor identity upon gauge fixing. However, one can introduce proper (additive) counterterms, order per order in perturbation theory, so that the Slavnov-Taylor identity is preserved, as should be the case for a non-anomalously quantized gauge theory.
(Of course, it is much easier to work with e.g.~dimensional regularization which is explicitly compatible with gauge invariance and does not allow mass divergences to begin with.)

Let us see how this enters the game in the current case.
If we do not set \(\delta(0)=0\), the variation \eqref{DeltaS} under the local shift \eqref{g2bis} with the identification \eqref{DeltaS2} yields
\begin{eqnarray}\label{variatie1}
\Delta S= -\int_{\Sigma_\pm} \dd[d-1]{x} \left[ b_i^{a,\pm}  n_\nu \frac{\delta F_{i\nu}^a}{\delta A_\rho^b}  \times n_\rho \omega^{\pm,b}\delta(0)\right]
=-\int_{\Sigma_\pm} \dd[d-1]{x} \left[ b_i^{a,\pm} D_i^{ab} \omega^{\pm,b}\right]\delta(0).
\end{eqnarray}
If the action $S$ were to contain a ``mass (counter)term''~(one that would be absent in dimensional regularization!)
\begin{eqnarray}\label{mass1}
 S_\text{count}=\int_{\Sigma_\pm} \dd[d-1]{x} \left[ b_i^{a,\pm} b_i^{a,\pm}\right]\frac{\delta(0)}{2},
\end{eqnarray}
the variation of this term would exactly cancel \eqref{variatie1}, so that \eqref{g2bis} indeed does generate a local invariance in the PMC case.

Now, the counterterm \eqref{mass1} is precisely the one needed to kill off the mass divergence entering the PMC effective action for the $b$-fields. Indeed, referring to \cite[Eq.~(21)]{Dudal2026}, there is a divergent term generated in the $1PI$ effective action
\begin{eqnarray}\label{mass2}
-\frac{1}{2}\int \ddk[d-1]{k} b_i^{a,\pm} b_i^{a,\pm}   \underbrace{\int \ddk{k_z}}_{=~\delta(0)}= -\frac{\delta(0)}{2}\int_{\Sigma_\pm} \dd[d-1]{x} b_i^{a,\pm} b_i^{a,\pm}=-S_\text{count}.
\end{eqnarray}
Said otherwise, when keeping the $\delta(0)$, one needs a proper counterterm to renormalize the theory, and the latter counterterm is exactly the action contribution needed to safeguard the symmetry generated by \eqref{g2bis}.

To end this section, we remind that in 4D, there also exists perfect electromagnetic conductor (PEMC) boundary conditions \cite{Rode:2017yqy,Schoger:2024zig,Dudal:2024PEMC} which are built from a linear combination of PMC and PEC\footnote{That is, mixing electric and magnetic field components.}, which in 4D Euclidean space read
\begin{equation}
    F_{\mu\nu}^a \cos\varphi + \frac{i}{2}\varepsilon_{\mu\nu\alpha\beta} F_{\alpha\beta}^a \sin\varphi=0,
\end{equation}
with duality angle \(\varphi\). In other dimensions, such combinations of the field strength tensor and its dual cannot be introduced, evidently. Using completely analogous arguments as those given for the PMC case, one can show that in the 4D PEMC case, one has the boundary gauge symmetry
\begin{eqnarray}
\Delta b_i^{a,\pm}(\cev x) = D_i^{ab}[A] \omega^{b,\pm}(\cev x),\qquad \Delta A_\mu^a= \cos\varphi^\pm \times\theta_\mu^{a,\pm}(\cev x)\delta_\pm
\end{eqnarray}
with duality angles \(\varphi^\pm\) per plate. Note however that, unlike in the Abelian case \cite{Rode:2017yqy,Dudal:2024PEMC}, one does not expect duality invariant physics in the non-Abelian case, that is, solely dependent on \(\varphi^+-\varphi^-\).

\subsection{Gauge fixing itself: PEC vs.~PMC}
The emergent gauge symmetries, \eqref{g2} vs.~\eqref{g2bis} have an important distinction: in the PEC case, the gauge field itself is left invariant, whilst in the PMC case it co-transforms.

In the PEC case, the emergent boundary gauge invariance \eqref{g2} is solely relying on the Bianchi identity, which is valid irrespective of equations of motion or the action itself. So the invariance holds even if we replace the classical Yang-Mills action with its (Landau) gauge fixed version, with or without the $d$-dimensional bulk Gribov--Zwanziger horizon term \cite{Zwanziger89,Vandersickel:2012tz} \footnote{Or even with the effective Curci-Ferrari mass term \cite{Pelaez21} in Landau gauge, $\frac{m^2}{2}\int \dd[d]{x} A_\mu^a A_\mu^a$, see \cite{Dudal2026}.}. Consequently, we must anyhow impose a second (boundary) gauge fixing to deal with the remaining (independent) local $\Delta$-invariance.

Let us now look at the PMC case. It is instructive to first consider the Landau gauge fixed version of the PMC action \eqref{action2},
\begin{eqnarray}\label{action2bis}
S'=\frac{1}{4}\int \dd[d]{x} F_{\mu\nu}^aF_{\mu\nu}^a + \int \dd[d]{x}\left[h^a \p_\mu A_\mu^a + \overline c^a \p_\mu D^{ab}_\mu c^b \right] +\int \dd[d]{x} \left[ b_i^{a,\pm}(\cev x) n_\nu F_{i\nu}^a\delta_\pm\right]
\end{eqnarray}
where the middle integral contains the standard Faddeev--Popov gauge fixing terms, with $h^a$ the Nakanishi--Lautrup multiplier enforcing the gauge condition and $c^a$, resp.~$\overline c^a$ the ghost, resp.~anti-ghost fields. This action is invariant under the BRST symmetry, generated by the nilpotent $s$,
\begin{eqnarray}\label{action2bis0}
s A_\mu^a=-D_\mu^{ab}c^b,\quad s c^a = \frac{g}{2}f^{abc}c^b c^c, \quad s\overline c^a = h^a,\quad sh^a=0.
\end{eqnarray}
At first sight, \eqref{g2bis} no longer define a symmetry as the FP gauge fixing terms are not invariant. Nonetheless, a boundary gauge fixing was still necessary in earlier work, as we already mentioned before, see \cite{Dudal2026,Dudal:2020yah}. Let us trace back the origin of this in more detail. It is easy to see that
\begin{eqnarray}\label{action2bis2}
\Delta S'=-\int \dd[d-1]{x} \left[\p_z h^a(\cev x, \pm L/2) \omega^{a,\pm}(\cev x) - \p_z \overline c^a(\cev x,\pm L/2) gf^{abc}\omega^{b,\pm}(\cev x)c^c(\cev x, \pm L/2)\right].
\end{eqnarray}
If we could assume that both $\p_z h^a$ and $\p_z \overline c^a$ vanish on the plates, the boundary gauge symmetry would be maintained even after the gauge fixing. To show that this is possible, let us introduce on the plates BRST field doublets according to $s\kappa^{a,\pm}(\cev x)=\chi^{a,\pm}(\cev x)$, $s\chi^{a,\pm}(\cev x)=0$ and replace \eqref{action2bis} as follows,
\begin{eqnarray}\label{action2bis3}
S' \to S' + \alpha\int \dd[d]{x}~s\left[\kappa^{a,\pm}(\cev x) \overline c^a \p_z\delta_\pm\right]= S'+ \alpha\int\dd[d-1]{x}\left[ \kappa^{a,\pm}(\cev x)\p_z\overline  c^a(\cev x, \pm L/2)+\chi^{a,\pm}(\cev x)\p_z h^a(\cev x, \pm L/2)\right].
\end{eqnarray}
For any non-zero choice of the parameter, this implements the desired assumptions. But since these extra assumptions are coming from a BRST cohomologically trivial modification of the action \cite{Piguet:1995er}, physical quantities will not depend on $\alpha$. Assuming $\alpha\neq 0$, if we set $\Delta \kappa^{a,\pm}$ and $\Delta \chi^{a,\pm}$ appropriately (in particular $\propto 1/\alpha$), then the $\Delta$-variation of \eqref{action2bis3} vanishes. Since the action also makes sense for $\alpha\to0$, including all expectation values computed with it, we can just as well consider this limit and conclude that, effectively, $\Delta S'=0$. In a more loose way of speaking, the ``breaking'' in \eqref{action2bis2} happens in a BRST trivial gauge fixing term, and as such, will not be ``visible'' to gauge invariant physics.

Said otherwise, the PMC action supplemented with (only) the standard Faddeev--Popov gauge fixing terms does respect the emergent boundary gauge symmetry.
On the other hand, if next to the Faddeeev--Popov terms, we were also to add e.g.~the bulk Gribov--Zwanziger horizon or a Curci--Ferrari mass term, this irrevocably breaks the shift symmetry \eqref{g2bis}. Indeed, the equations of motion of the $A_\mu^a$-field drastically change due to the modified action, but the variation of the boundary condition term remain proportional to the ``standard YM'' equations of motion, hence the mismatch and loss of symmetry. The breaking terms are no longer BRST trivial, hence they become physically relevant.

In \cite{Dudal:2026sue}, the bulk generalization of the $d=4$ Gribov-Zwanziger horizon action was studied at leading order in the presence of two parallel plates with separation $L$.  This concerns the ``standard'' bulk Landau gauge. It was found that, although in the intermediate steps, the plates are present, at the end the $L$ drops out from the corresponding $d=4$ Gribov gap equation as the additional contributions are suppressed with a $V_3/V_4\to0$ factor where $V_3$ is the (infinite) plate volume and $V_4$ the (infinite) higher-dimensional bulk volume. 

These computations, including that of the Casimir energy, required an explicit gauge fixing of the emergent boundary invariance for the PEC case. The Gribov copies related to the emergent $d=3$ local gauge invariance were ignored in that first approximation. 

However, speculating now which new ingredient the latter might bring about: assuming Landau gauge for the emergent gauge invariance as well, a $d=3$ Gribov-Zwanziger reasoning will lead to (some form of the) $d=3$ Gribov-Zwanziger actions on the plates, with their own (additional) $d=3$ Gribov masses and gap equations, now explicitly $L$-dependent. It is hence our expectation that this can give a an additional non-perturbative contribution  to the Yang-Mills Casimir energy/force, at least in the PEC case.

In the PMC case, the bulk Gribov-Zwanziger action already removes the emergent boundary gauge invariance, and indeed in \cite{Dudal:2026sue} no extra gauge fixing was required to carry out the computations, unlike in the PEC or (standard) perturbative Yang-Mills case. A consequence is that the Gribov mass to zero limit of the PMC analysis does not match with the same PEC limit, that is, the perturbative result. This is exactly what we called a van Dam--Veltman--Zakharov-like discontinuity, already noticed before in \cite{Dudal2026}, see also \cite{Karabali:2025olx}. In the latter reference, a proposal was made to restore the smooth ``mass~$\to 0$'' by redefining PMC, see their eq.(50). At first sight, their new term would have to contain a covariant derivative, and the gauge variation then a (non-vanishing) commutator of two covariant derivatives, thereby hampering the gauge invariance. We do not regard PMC boundary conditions as inconsistent or intrinsically unacceptable. Rather, within the present Gribov–Zwanziger/Curci–Ferrari treatment, standard PMC boundary conditions define a boundary problem whose vanishing-mass limit is not continuously connected to the perturbative PEC/Maxwell result. As said, an Abelian prescription restoring a smooth limit has been proposed in Ref. \cite{Karabali:2025olx}, but a gauge-covariant non-Abelian generalization is presently unknown. Clarifying whether such a generalization exists requires a separate investigation beyond the scope of the present work.

\section{Conclusion and Outlook}

We have devised a way to construct explicit zero-modes\footnote{In fact infinitely many, given the freedom in \eqref{free1}.} of the Faddeev-Popov operator in Yang-Mills theory in the presence of PEC and/or PMC plates.
The boundary conditions of the plates restrict the allowed gauge field configurations, but nonetheless, a generalization of the Henyey procedure of \cite{Henyey:1978qd,Capri:2012ev} still yields examples of zero-modes of the Faddeev-Popov operator under these additional restrictions. We have constructed examples with both axial and spherical symmetry, where the latter class was quite useful to find gauge field configurations with finite support between the plates so that the boundary constraints trivialize. This procedure also allows to construct both zero-modes and gauge fields with finite $L_2$-norm.

The mere existence of these zero-modes implies that the Gribov ambiguity still plagues non-Abelian gauge theories in presence of PEC/PMC plates. 

When quantizing the theory, one thus still needs to deal with these Gribov copies in one way or another. In future work, we will focus on the Gribov-Zwanziger formalism \cite{Gribov78,Zwanziger89}.
However, we showed that an extra complication shows up: as was noted in previous works in the Abelian approximation, if one lifts PEC or PMC boundary conditions into the action using multiplier fields, a new boundary gauge symmetry of the form \(b_i \rightarrow b_i + \p_i\omega(\cev x)\) emerges. In this work, we have discussed this symmetry in more detail and at the non-Abelian level: we made explicit the corresponding transformations of the gauge and boundary fields, and carefully showed that the action is indeed invariant, both using dimensional regularization and an UV cut-off with a mass counterterm. It is vital to take into account both these gauge symmetries, since each symmetry will give rise to its own further restriction of allowed gauge and boundary field configurations. The precise consequences of these boundary gauge invariances in a GZ setup will be part of forthcoming work, including the difference between the PEC and PMC gauge fixing cases. In particular, although we do not regard PMC boundary conditions as intrinsically inconsistent, their vanishing-mass limit is not continuously connected to the perturbative PEC/Maxwell result within the present treatment. Whether a gauge-covariant non-Abelian prescription can restore a smooth limit remains an open question.

\section*{Acknowledgments}
D.D., L.R.~and S.S.~are grateful for the hospitality at CECs and USS, where the main ideas in this paper took first shape. We thank F.~Canfora for interesting comments. We are grateful to the authors of \cite{Karabali:2025olx} and \cite{Vercauteren:2026kdz} for discussions that motivated the second part of this work.
The work of D.D.~was supported by KU Leuven IF project C14/21/087. L.R.~acknowledges the Ministero dell’Universit\`a e della Ricerca (MUR), PRIN2022 program (Grant PANTHEON 2022E2J4RK) for partial support. The work of S.S.~was funded by FWO PhD-fellowship fundamental research (file number: 1132823N).

\bibliography{allpapers}

\end{document}